\documentclass[article,preprint,showpacs,amsfonts]{revtex4-1}
\usepackage{indentfirst}
\usepackage{amsmath,amssymb,amsthm}
\usepackage{graphicx}
\usepackage{epsfig}
\usepackage{epstopdf}
\def\ra{\rangle}
\def\la{\langle}

\usepackage{enumerate}

\begin{document}

\title{The quantum uncertainty relations of quantum channels}
\smallskip
\author{Shi-Yun Kong$^1$}
\author{Ming-Jing Zhao$^2$}
\thanks{Corresponding author: zhaomingjingde@126.com}
\author{Zhi-Xi Wang$^1$}
\thanks{Corresponding author: wangzhx@cnu.edu.cn}
\author{Shao-Ming Fei$^{1,3}$}
\thanks{Corresponding author: feishm@cnu.edu.cn}
\affiliation{{\footnotesize $^1$School of Mathematical Sciences, Capital Normal University, Beijing 100048, China}\\
{\footnotesize $^2$School of Science, Beijing Information Science and Technology University, Beijing 102206, China}\\
{\footnotesize $^3$Max-Planck-Institute for Mathematics in the Sciences, 04103 Leipzig, Germany}
}

\begin{abstract}
The uncertainty relation reveals the intrinsic difference between the classical world and the quantum world. We investigate the quantum uncertainty relation of quantum channel in qubit systems. Under two general measurement bases, we first derive the quantum uncertainty relation for quantum channels with respect to the relative entropy of coherence. Then we obtain the quantum uncertainty relation for unitary channels with respect to the $l_1$ norm of coherence. Some examples are given in detail.

\end{abstract}

\maketitle
\section{Introduction}

The Heisenberg's uncertainty principle \cite{heisenberg} is one of the most fundamental features in quantum world, which is the essential difference from classical world. Given two incompatible measurement, the uncertainty relation shows that the outcomes cannot be accurately predicted simultaneously. The Heisenberg-Robertson uncertainty relation \cite{robertson} provides the bounds of two observables based on variance.
Deutsch \cite{deutsch} establishes the celebrated entropic uncertainty relation.
Afterward, Maassen and Uffink \cite{maassen} further propose the state-independent uncertainty relation based on the Shannon entropy of measurement outcomes.
The uncertainty relation for general quantum
channels is also investigated \cite{fu,zhou}.
As for applications, the uncertainty relations have played significant roles in the realm of quantum information processing, including the quantum key distribution \cite{vallone,coles}, entanglement witnesses \cite{berta}, quantum steering \cite{walborn}, quantum teleportation \cite{hu1}, and quantum metrology \cite{Giovannetti}.

In general, the nominal uncertainty of a measurement comprises classical (predictable) and quantum (unpredictable) components arising from classical noise and quantum effects \cite{yuan1}. Luo $et\ al.$ shows that the quantum uncertainty is equivalent to quantum coherence \cite{luo}, which is viewed as an essential resource \cite{marvian,wu,streltsov}. The development of the quantum coherence framework for quantum states \cite{baumgratz} has attracted a growing interest in quantifying the coherence, such as the $l_1$ norm of coherence \cite{baumgratz}, the relative entropy of coherence \cite{baumgratz}, the intrinsic randomness of coherence \cite{yuan1}, coherence concurrence \cite{qi}, the robustness of coherence\cite{carmine} and the fidelity-based coherence \cite{liu1} and so on.

Since the quantum coherence depends on the measurement basis, it is interesting to study the constraints of the quantum coherence on different measurement bases \cite{luo}, which is the quantum uncertainty relation. For the qubit states, Yuan $et\ al.$ \cite{yuan} derived the quantum uncertainty relation for qubit states on two measurement bases for the relative entropy of coherence, the $l_1$ norm of coherence, and the coherence of formation respectively. In high dimensional systems,
the quantum uncertainty relations of Tsallis relative entropy of coherence and R\'{e}nyi relative entropy of coherence are derived \cite{Rastegin2023,FuGang Zhang},
which are then generalized to the sandwiched R\'{e}nyi relative entropy of coherence and unified ($\alpha$,$\beta$)-relative entropy of coherence
\cite{Haijin Mu}. The quantum uncertainty relations for coherence measure based on the Wigner-Yanase skew information are also
established \cite{Shun-Long Luo}. In \cite{hu}, the quantum uncertainty relations of the geometric coherence on two and three general measurement bases are derived.

Analogous to quantum states, Xu established a framework for quantifying the coherence of quantum channels \cite{xu}.
Subsequently, the study of the coherence of quantum channels has attracted more and more attention,
such as the coherence based on skew information for quantum channels \cite{xuan}, the coherence based on trace distance for quantum channels \cite{fan},
and maximum relative entropy of coherence \cite{jin}.  Moreover, the coherence of quantum channels is also related to the measurement basis. %

In this work, we investigate the quantum uncertainty relation of quantum channels in qubit systems. In section II, we introduce the coherence framework and some basic concepts needed in this paper. In section III, for any given two measurement bases, we first derive the quantum uncertainty relation of quantum channels in terms of the relative entropy of coherence, and then we obtain the quantum uncertainty relation for unitary channels in terms of the $l_1$ norm of coherence. Detailed examples are presented to illustrate these results. We conclude in section IV.

\maketitle
\section{Preliminaries}
Let $H_A$ be a Hilbert space with dimension $|A|$, and $\mathbb{J}=\{|j\ra\}$ be a orthonormal basis of $H_A$. Denote $\mathcal D_A$ the set of all density operators on $H_A$. The incoherent states are quantum states that are diagonal under the reference basis $\mathbb{J}$, that is $\rho=\sum_j p_j|j\ra\la j|$, where $p_j\geqslant0, \sum p_j=1$. Otherwise, it is called coherent. The set of all the incoherent states is denoted as $\mathcal{I}_\mathbb{J}$.

Under the reference basis $\mathbb{J}=\{|j\ra\}$, for any given mixed state $\rho=\sum_{i,j} \rho_{ij} |i\rangle\langle j|$, the relative entropy of coherence is defined as
\begin{equation}
C_{\rm{rel}}^{\mathbb{J}}(\rho)=\underset{\sigma\in\mathcal{I}_\mathbb{J}}{\min}S(\rho\|\sigma)=S({\rm{diag}}^\mathbb{J} (\rho))-S(\rho),
\end{equation}
where $S(\rho\|\sigma)=\rm{Tr}(\rho \rm{log}\rho-\rho\rm{log}\sigma)$ is the quantum relative entropy, 
$S(\rho)=-\rm{Tr}(\rho\rm{log}\rho)$ is the von Neumann entropy and ${\rm{diag}}^\mathbb{J}(\rho)=\sum_j \rho_{jj}|j\ra\la j|$ is the state of $\rho$ 
after a completely dephasing channel in the basis $\mathbb{J}$.
The $l_1$ norm of coherence of $\rho$ is defined as
\begin{equation}C_{l_1}^{\mathbb{J}}(\rho)=\underset{i\neq j}{\sum}|\rho_{ij}|.
\end{equation}
These two coherence measures are widely adopted and applied in quantum tasks such as quantum assistance \cite{zhao} and quantum coherence distribution \cite{machado}.

A quantum channel $\Phi$ is a linear completely positive and trace preserving (CPTP) map, which can be represented by the Kraus operators $\Phi=\{M_m\}$, with $\sum_m M_m^\dagger M_m=I$. Moreover, a quantum channel can also be represented by the Choi matrix
\begin{equation}
J_\Phi=\sum_{j,k}|j\ra\la k|\otimes\Phi(|j\ra\la k|)=\sum_{m,j,k}|j\ra\la k|\otimes M_m|j\ra\la k|M_m^\dagger.
\end{equation}
Let $\mathcal C_{AB}$ be the set of all quantum channels from $\mathcal D_A$ to $\mathcal D_B$.
In the framework of coherence theory of quantum channels,
$\Phi\in\mathcal{C}_{AB}$ is called an incoherent quantum channel if $\Upsilon(\Phi)=\Phi$, where $\Upsilon(\Phi)=\Delta_B \Phi \Delta_A$, $\Delta_A$ and $\Delta_B$ are resource destroying maps \cite{liu}. Or else the quantum channel is called coherent. In order to quantify the coherence of quantum channels,
a coherence measure $C$  is proposed which  should satisfy the following conditions:
(i) Positivity: For any quantum channel $\Phi$, it has $C(\Phi)\geqslant0$; $C(\phi)=0$ if and only if $\Phi$ is an incoherent channel;
(ii) Monotonicity: $C(\Phi)$ cannot increase under incoherent maps;
(iii) Convexity: $C(\Phi)$ is convex.
 In \cite{xu}, the author introduced a resource theory for quantifying the coherence of quantum channels by using the coherence measure for quantum states.
Suppose $C$ is any coherence measure for quantum states, then the coherence of quantum channel $\Phi$ can be characterized by the Choi matrix $J_\Phi$ as \cite{xu}
\begin{equation}\label{eq c def}
C(\Phi)=C(\frac{J_\Phi}{|A|}).
\end{equation}
Here $\frac{J_\Phi}{|A|}$ is factually a density matrix. So the coherence of quantum channel is quantified by that of the corresponding quantum state.

We aim to get the quantum uncertainty relations for quantum channels in qubit systems. Since the quantum uncertainty relations depends on the measurement basis, so we expect to describe it by the functions follows:
 \begin{equation}
 C^\mathbb{X}(\Phi)+C^\mathbb{Z}(\Phi)\geqslant f(\mathbb{X},\mathbb{Z},\Phi),
 \end{equation}
 where $\mathbb{X}$ and $\mathbb{Z}$ are two general measurement bases, $C$ is a coherence measure for quantum channels, and $f$ is a function of the measurement bases and the quantum channel $\Phi$.

\section{The quantum uncertainty relation of quantum channels with respect to the relative entropy of coherence}

In this section, we aim to derive the quantum uncertainty relation of quantum channels in qubit systems with respect to the relative entropy of coherence.
For any two orthonormal bases $\mathbb{X}$ and $\mathbb{Z}$, we denote $c_{\min}={\min}_{|x\ra\in\mathbb{X},\ |z\ra\in\mathbb{Z}}|\la x|z\ra|^2$ as their incompatibility. In the meantime, we denote
\begin{eqnarray}\label{eq c}
c_{\max}={\max}_{|x\ra\in\mathbb{X},\ |z\ra\in\mathbb{Z}}|\la x|z\ra|^2
\end{eqnarray}
as the maximum overlap of the two measurements. In qubit systems, it has $c_{\max}=1-c_{\min}$. For convenience, we use $c_{\max}$ throughout this paper to characterize the quantum uncertainty relation of quantum channels.
Before the study of the quantum uncertainty relation, we need the following lemma first.

{\bf Lemma 1.} In $d$-dimensional systems, for any two normalized vectors $|x\ra,|z\ra$ and a density matrix $A$  in form of $A=|r_1\ra\la r_1|+\dots+|r_n\ra\la r_n|$, with $\la r_1|r_1\ra+\dots+\la r_n|r_n\ra=1$. Denote $\la x|A|x\ra=a,\ \la z|A|z\ra=b$ and $|\la x|z\ra|^2=c$, then we have $a+b\leqslant1+\sqrt{c}$ and $|a-b|\leqslant\sqrt{1-c}$. When $d=2$, we also have $1-\sqrt c\leqslant a+b.$

{\bf Proof.}
First, if the density matrix $A$ is rank one, then these inequalities are obviously true as proved in \cite{yuan}. Now we prove the inequality $a+b\leqslant1+\sqrt{c}$
for general density matrix $A$ in form of $A=\sum_i|r_i\ra\la r_i|$.
For each vector $|r_i\ra,\ i=1,\cdots, n$, it has
\begin{eqnarray*}
\frac{\la x|r_i\ra\la r_i|x\ra}{\la r_i|r_i\ra}+\frac{\la z|r_i\ra\la r_i|z\ra}{\la r_i|r_i\ra}\leqslant1+\sqrt{c},
\end{eqnarray*}
where we have used the inequality $a+b\leqslant1+\sqrt{c}$ for rank one $A=\frac{|r_i\ra\la r_i|}{\la r_i|r_i\ra}$.
Therefore we can get
\begin{eqnarray*}
\la x|A|x\ra + \la z|A|z\ra=\sum_{i}\la x|r_i\ra\la r_i|x\ra+\sum_{i}\la z|r_i\ra\la r_i|z\ra\leqslant\sum_{i}\la r_i|r_i\ra(1+\sqrt{c})=1+\sqrt{c},
\end{eqnarray*}
which implies $a+b\leqslant1+\sqrt c$. The inequalities $|a-b|\leqslant\sqrt{1-c}$ and $1-\sqrt c\leqslant a+b$ can be derived similarly.\qed

Now we are ready to prove the quantum uncertainty relation for quantum channels with respect to the relative entropy of coherence.

{\bf Theorem 1.} For any quantum channel $\Phi$ and any two measurement bases $\mathbb{X}=\{|x\ra,\ |x_\perp\ra\}$ and $\mathbb{Z}=\{|z\ra,\ |z_\perp\ra\}$, the quantum uncertainty relation of quantum channels with respect to the relative entropy of coherence is
\begin{equation}\label{eq rel}
C_{\rm rel}^{\mathbb{X}}(\Phi)+C_{\rm rel}^{\mathbb{Z}}(\Phi)\geqslant H(\sqrt {c_{\max}})-2S(\frac {J_\Phi}{2})+2{\rm log}2
\end{equation}
with $c_{\max}$ in Eq. (\ref{eq c}) and ${J_\Phi}$ is the Choi matrix of the quantum channel $\Phi$.

{\bf Proof.}
Note that in qubit systems,
the Choi matrix of a quantum channel $\Phi$ is factually a $4\times 4$ matrix as
\begin{align}\begin{split}
J_{\Phi}=\sum_{j,k=0}^{1}|j\ra\la k|\otimes\Phi(|j\ra\la k|)=
\begin{pmatrix}\Phi(|0\ra\la0|)&\Phi(|0\ra\la1|)\\
\Phi(|1\ra\la0|)&\Phi(|1\ra\la1|)\end{pmatrix}.
\end{split}\end{align}
First under the orthonormal basis $\mathbb{X}$, the relative entropy of coherence of the quantum channel is
\begin{equation}C_{\rm rel}^{\mathbb{X}}(\Phi)=S({\rm{diag}}^\mathbb{X}(\frac{J_\Phi}{2}))-S(\frac{J_{\Phi}}{2}).\end{equation}
Suppose $\la x|\sum_m M_m|0\ra\la0|M_m^\dagger|x\ra=a_1,\la x|\sum_m M_m|1\ra\la1|M^\dagger_m|x\ra=a_2$. Then $\la x_\perp|\sum_{m}M_{m}|0\ra\\\la0|M_m^{\dagger}|x_\perp\ra=1-a_1,\la x_\perp|\sum_{m}M_{m}|1\ra\la1|M_m^{\dagger}|x_\perp\ra=1-a_2.$ In this way, the diagonal matrix of $\frac{J_{\Phi}}{2}$ is
\begin{align}\begin{split}{\rm diag}^{\mathbb{X}}(\frac{J_{\Phi}}{2})=\begin{pmatrix}\frac{a_1}{2}&\ &\ &\ \\\ &\frac{1-a_1}{2}&\ &\ \\\ &\ &\frac{a_2}{2}&\ \\\ &\ &\ &\frac{1-a_2}{2}\end{pmatrix}.
\end{split}\end{align}
This implies that
\begin{eqnarray*}
S({\rm{diag}}^{\mathbb{X}}(\frac{J_{\Phi}}{2}))&=&-\frac{a_1}{2}{\rm log}\frac{a_1}{2}-\frac{1-a_1}{2}{\rm log}\frac{1-a_1}{2}-\frac{a_2}{2}{\rm log}\frac{a_2}{2}-\frac{1-a_2}{2}{\rm log}\frac{1-a_2}{2}\\
&=&\frac{1}{2}(H(a_1)+H(a_2))+{\rm log}2,
\end{eqnarray*}
with $H(x)=-x{\rm log}x-(1-x){\rm log}(1-x)$ the binary entropy.

Under the orthonormal basis $\mathbb{Z}$, we assume $\la z|\sum_{m}M_m|0\ra\la0|M_m^\dagger|z\ra=b_1$ and $\la z|\sum_m M_m|1\ra\la1|M_m^\dagger|z\ra=b_2$, then
we have similarly $S({\rm diag}^{\mathbb{Z}}(\frac{J_{\Phi}}{2}))=\frac{1}{2}(H(b_1)+H(b_2))+{\rm log}2$. Therefore the sum
is
\begin{eqnarray}\label{eq c rel}
S({\rm{diag}}^{\mathbb{X}}(\frac{J_{\Phi}}{2})) +S({\rm{diag}}^{\mathbb{Z}}(\frac{J_{\Phi}}{2}))&=&\frac{1}{2}(H(a_1)+H(b_1))+\frac{1}{2}(H(a_2)+H(b_2))+2{\rm log}2.
\end{eqnarray}
Let $g=S({\rm{diag}}^{\mathbb{X}}(\frac{J_{\Phi}}{2})) +S({\rm{diag}}^{\mathbb{Z}}(\frac{J_{\Phi}}{2}))$ and
the binary function $f(x,y)=\frac{1}{2}(H(x)+H(y))$, then Eq. (\ref{eq c rel}) above can be reduced to
\begin{eqnarray}
 g(a_1,a_2,b_1,b_2)=f(a_1,b_1)+f(a_2,b_2)+2{\rm log}2.
\end{eqnarray}
Next we aim to find the minimal value of $g(a_1,a_2,b_1,b_2)$. Firstly, $f(x,y)$ has second-order continuous partial derivatives and
$\frac{\partial{f(x,y)}}{\partial{{x}}}|_{(\frac{1}{2},\frac{1}{2})}=\frac{\partial{f(x,y)}}{\partial{{y}}}|_{(\frac{1}{2},\frac{1}{2})}=0$,
then $P_0=(\frac{1}{2},\frac{1}{2})$ is the unique stationary point of $f(x,y)$.
Since binary entropy $H(x)$ is concave, i.e. $H''(x)\leqslant0$, the Hesse matrix of $f(x,y)$ at $P_0$ is a negative definite matrix. According to the sufficient condition of extreme value, $f$ has the maximal value of 1 at $P_0$. Let $A_i=a_i+b_i, B_i=b_i-a_i,i=1,2$. According to Lemma 1, $A_i\in [1-\sqrt {c_{\max}}, 1+\sqrt{c_{\max}}]$ and $B_i\in[0,\sqrt{1-c_{\max}}]$. Here, without loss of generality, we assume $a_i\leqslant b_i$. Then the function $g$ can be expressed as $g=\frac{1}{2}f(\frac{A_1-B_1}{2},\frac{A_1+B_1}{2})+\frac{1}{2}f(\frac{A_2-B_2}{2},\frac{A_2+B_2}{2})+2{\log}2$. Taking derivative $g$ on $A_i$ and $B_i$ yields that
\begin{equation}
\frac{\partial g}{\partial A_i}=\frac{1}{2}\frac{\partial f}{\partial A_i}=(H'(\frac{A_i-B_i}{2})+H'(\frac{A_i+B_i}{2}))=\frac{1}{4}(H'(a_i)+H'(b_i)),
\end{equation}
and
\begin{equation}
\frac{\partial g}{\partial B_i}=\frac{1}{2}\frac{\partial f}{\partial B_i}=\frac{1}{4}(-H'(\frac{A_i-B_i}{2})+H'(\frac{A_i+B_i}{2}))=\frac{1}{4}(H'(b_i)-H'(a_i)),
\end{equation}
respectively.
Furthermore, the second-order derivatives of $g$ with respect to $A_i$ and $B_i$ are
\begin{equation}
\frac{\partial^2 g}{\partial {A_i}^2}=\frac{\partial^2 g}{\partial {B_i}^2}=\frac{1}{8}(H''(\frac{A_i-B_i}{2})+H''(\frac{A_i+B_i}{2}))=\frac{1}{8}(H''(a_i)+H''(b_i)).
\end{equation}
It is obvious that $\frac{\partial^2 g}{\partial {A_i}^2}=\frac{\partial^2 g}{\partial {B_i}^2}\leqslant0$ and $\frac{\partial g}{\partial B_i}\leqslant0$.
According to $f(\frac{A_i-B_i}{2},\frac{A_i+B_i}{2})=f(1-\frac{A_i-B_i}{2},1-\frac{A_i+B_i}{2})$, so $g$ is symmetric about $A_i=1$. In the following, we only need to consider the case $A_i\leqslant1$. $\frac{\partial g}{\partial B_i}\leqslant0$ shows that $g$ decreases monotonously with respect to $B_i$. To find the minimum value of $g$, we should make $A_i$ as small as possible and $B_i$ as large as possible. The values of $A_i$ and $B_i$ have the following two situations:\\
(i) $A_i=B_i=1-\sqrt{ c_{\rm max}}$, and $g=H(\sqrt {c_{\rm max}})+2{\rm log}2;$\\
(ii) $A_i=B_i=\sqrt{1-c_{\rm max}}$, and $g=H(\sqrt{1-c_{\rm max}})+2{\rm log}2$.\\
In light of $|\sqrt{ c_{\rm max}}-\frac{1}{2}|\geqslant|\sqrt{1-c_{\rm max}}-\frac{1}{2}|$ by $c_{\rm max}\in[\frac{1}{2},1]$,  and $H(x)$ is symmetric about $x=\frac{1}{2}$, we derive that $H({\sqrt{ c_{\rm max}}})\leqslant H({\sqrt{1-c_{\rm max}}})$. As a result, we can get the minimum value of $g$ is $ H(\sqrt{ c_{\rm max}})+2{\rm log}2$, which is equivalent to
\begin{eqnarray*}
S({\rm{diag}}^{\mathbb{X}}(\frac{J_{\Phi}}{2})) +S({\rm{diag}}^{\mathbb{Z}}(\frac{J_{\Phi}}{2}))\geqslant H(\sqrt{ c_{\rm max}})+2{\rm log}2.
\end{eqnarray*}
Thus, we get the quantum uncertainty relation of quantum channels with respect to the relative entropy of coherence in Eq. (\ref{eq rel}).
\hfill$\square$

The quantum uncertainty relation in
Theorem 1 demonstrates the constraint of the coherence of quantum channels under two orthonormal bases by the incompatibility of the two orthonormal bases and the entropy of the Choi matrices of the quantum channels. In general, for two measurement bases $\mathbb{X}$ and $\mathbb{Y}$, if a quantum channel $\Phi$ with Kraus operators $\{M_m\}$ satisfies the following condition:
\begin{equation}\sum_m|\la x|M_m|i\ra|^2+\sum_m|\la y|M_m|i\ra|^2=\sum_m|\la x|M_m|i\ra|^2-\sum_m|\la y|M_m|i\ra|^2=1-\sqrt{c_{\rm max}},\end{equation}
for some $|x\ra\in\mathbb{X},\ |y\ra\in\mathbb{Y},\ i=0,1$, then the quantum uncertainty relation of a quantum channel reaches the lower bound $H(\sqrt{c_{\rm max}})-2S(\frac{J_\Phi}{2})+2{\rm log}2$, i.e., the equality in (\ref{eq rel}) holds.
Now we illustrate the quantum uncertainty relation of quantum channels by three specific examples.

{\bf Example 1.} Firstly we consider the bit flip channel as an example. The bit flip channel $\Phi$ has Kraus operators
\begin{equation}M_1=\sqrt pI=\sqrt p\begin{pmatrix}1&0\\0&1\end{pmatrix},\ M_2=\sqrt{1-p}X=\sqrt{1-p}\begin{pmatrix}0&1\\1&0\end{pmatrix}.\end{equation}
Given two measurement bases $\mathbb{X}=\{|0\ra,\ |1\ra\}$ and $\mathbb{Z}=\{|+\ra,\ |-\ra\}$ with $\{|+\rangle=\frac{1}{\sqrt{2}}(|0\ra+|1\ra)$, $|-\rangle=\ \frac{1}{\sqrt{2}}(|0\ra-|1\ra)\}$, $c_{\rm max}=\frac{1}{2}$. In this case we get that the Choi matrices of the bit flip channel under these two orthonormal bases are
\begin{align}J_\Phi^\mathbb{X}=\begin{pmatrix}p&\ &\ &p\\\ &1-p&1-p&\ \\\ &1-p&1-p&\ \\p&\ &\ &p\end{pmatrix},\ {\rm and}\ J_\Phi^\mathbb{Z}=\begin{pmatrix}\frac{1}{2}&\frac{1}{2}(2p-1)&\frac{1}{2}&\frac{1}{2}(1-2p)\\\frac{1}{2}(2p-1)&\frac{1}{2}&\frac{1}{2}(2p-1)&-\frac{1}{2}\\
\frac{1}{2}&\frac{1}{2}(2p-1)&\frac{1}{2}&\frac{1}{2}(1-2p)\\\frac{1}{2}(1-2p)&-\frac{1}{2}&\frac{1}{2}(1-2p)&\frac{1}{2}\end{pmatrix},\end{align}
respectively.
By calculation we can derive that $S({\rm diag}^\mathbb{X}(\frac{J_\Phi}{2}))=H(p)+{\rm log}2,\ S({\rm diag}^\mathbb{Z}(\frac{J_\Phi}{2}))=2{\rm log}2$, and $S(\frac{J_\Phi}{2})=H(p)$ respectively. So
the sum of the relative entropy coherences of the bit flip channel under these two orthonormal bases is
$C_{\rm rel}^\mathbb{X}(\Phi)+C_{\rm rel}^\mathbb{Z}(\Phi)=3{\rm log}2-H(p) $. By Theorem 1, the lower bound of the quantum uncertainty relation is $C_{\rm rel}^\mathbb{X}(\Phi)+C_{\rm rel}^\mathbb{Z}(\Phi)\geqslant H(\frac{1}{\sqrt{2}})-2H(p)+2{\rm log}2.$ The comparison between the sum of the relative entropy coherences and the lower bound is shown in Fig.1.

\begin{figure}[h]
\includegraphics{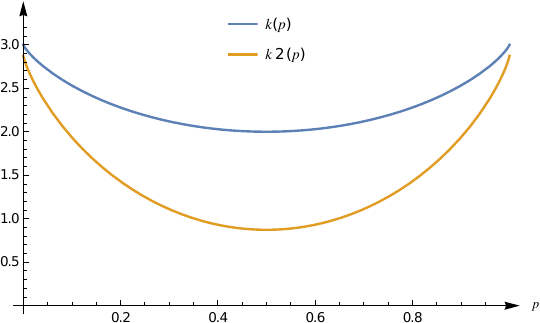}
\caption {The quantum uncertainty relations of bit flip channel with respect to the relative entropy coherence. $k(p)$ is the sum of the relative entropy coherence of the quantum channel with measurement bases $\mathbb{X}$ and $\mathbb{Z}$. $k2(p)$ is the lower bound of the quantum uncertainty relation at the right hand side of (\ref{eq rel}).}
\end{figure}

{\bf Example 2.} Secondly we consider a unitary channel $\sigma_x=|0\rangle\langle1|+ |1\rangle\langle0|$. Given two measurement bases, $\mathbb{X}=\{|0\ra, |1\ra\}$ and $\mathbb{Y}=\{|y_1\ra,|y_2\ra\}$ with $|y_1\ra=(1-\frac{3-\sqrt5}{2})^{\frac{1}{2}} |0\ra+ (\frac{3-\sqrt5}{2})^{\frac{1}{2}}|1\ra$, $|y_2\ra=-(\frac{3-\sqrt5}{2})^{\frac{1}{2}} |0\ra + (1-\frac{3-\sqrt5}{2})^{\frac{1}{2}} |1\ra$ with $c_{\max}=\frac{3-\sqrt5}{2}$.
We derive that $S({\rm diag}^\mathbb{X}(\frac{J_{\sigma_x}}{2}))
=\log2$, $S({\rm diag}^\mathbb{Y}(\frac{J_{\sigma_x}}{2}))=H(\frac{3-\sqrt5}{2})+\log2$ and $S(\frac{J_{\sigma_x}}{2})=1\log1=0$, respectively. Then we can get the sum of the relative entropy coherence of $\sigma_x$ under two orthonormal bases $\mathbb{X}$ and $\mathbb{Y}$ is $C^\mathbb{X}_{\rm rel}(\sigma_x)+C^\mathbb{Y}_{\rm rel}(\sigma_x)=H(\frac{3-\sqrt5}{2})+2\log2$, which is equal to the right hand side of (\ref{eq rel}), $H(\sqrt{c_{\max}})-2S(\frac{J_{\sigma_x}}{2})+2\log2$.

{\bf Example 3.} Thirdly we consider a phase damping channel $\Phi$ with Kraus operators \begin{equation}E_0=\begin{pmatrix}1&0\\0&\sqrt{1-\lambda}\end{pmatrix}\ {\rm and}\ E_1=\begin{pmatrix}0&0\\0&\sqrt{\lambda}\end{pmatrix}.\end{equation}
If we choose two measurement bases as $\mathbb{X}=\{|0\ra,|1\ra\}$ and $\mathbb{Z}=\{|+\ra,|-\ra\}$ with $c_{\rm max}=1/2.$ Then the Choi matrix of the phase damping channel under bases $\mathbb{X}$ and $\mathbb{Z}$ are
\begin{equation}
J_\Phi^{\mathbb {X}}=\begin{pmatrix}1&0&0&\sqrt{1-\lambda}\\
0&0&0&0\\0&0&0&0\\\sqrt{1-\lambda}&0&0&1\end{pmatrix}\ {\rm and}\ J_\Phi^{\mathbb{Z}}=\frac{1}{2}\begin{pmatrix}
1&1&\sqrt{1-\lambda}&-\sqrt{1-\lambda}\\
1&1&\sqrt{1-\lambda}&-\sqrt{1-\lambda}\\
\sqrt{1-\lambda}&\sqrt{1-\lambda}&1&1\\
-\sqrt{1-\lambda}&-\sqrt{1-\lambda}&1&1\end{pmatrix},
\end{equation}
respectively. By calculation we can derive that $S({\rm diag}^{\mathbb{X}}(\frac{J_\Phi}{2}))={\rm log}2,\ S({\rm diag}^{\mathbb{Z}}(\frac{J_\Phi}{2}))=2{\rm log}2$, and $S(\frac{J_\Phi}{2})=H(\frac{1+\sqrt{1-\lambda}}{2})$ respectively. So the sum of the relative entropy of coherences of the phase damping channel under two orthonormal bases is $C_{\rm rel}^{\mathbb{X}}(\Phi)+C_{\rm rel}^{\mathbb{Z}}(\Phi)=3{\rm log}2-2H(\frac{1+\sqrt{1-\lambda}}{2})$. By Theorem 1, the lower bound of the quantum uncertainty relation is $C_{\rm rel}^{\mathbb{X}}(\Phi)+C_{\rm rel}^{\mathbb{Z}}(\Phi)\geqslant H(\frac{1}{\sqrt{2}})-2H(\frac{1+\sqrt{1-\lambda}}{2})+2{\rm log}2$.

If we choose two measurement bases as $\mathbb{X}=\{|0\ra,|1\ra\}$ and $\mathbb{Y'}=\{|y'_1\ra,|y'_2\ra\}$ with $|y'_1\ra=\frac{\sqrt3+\sqrt6{\rm i}}{4}|0\ra+\frac{\sqrt{2}+\sqrt{5}{\rm i}}{4}|1\ra,\ |y'_2\ra=\frac{\sqrt2-\sqrt5{\rm i}}{4}|0\ra+\frac{-\sqrt3+\sqrt6{\rm i}}{4}|1\ra$ with $c_{\rm max}=9/16$. We derive that $S({\rm diag}^{\mathbb{X}}(\frac{J_\Phi}{2}))={\rm log}2$, $S({\rm diag}^{\mathbb{Y'}}(\frac{J_\Phi}{2}))=H(\frac{9}{16})+{\rm log}2$ and $S(\frac{J_\Phi}{2})=H(\frac{1+\sqrt{1-\lambda}}{2})$, respectively. Then we can get the sum of the relative entropy coherence of $\Phi$ under two orthonormal bases $\mathbb{X}$ and $\mathbb{Y'}$ is $C_{\rm rel}^{\mathbb{X}}(\Phi)+C_{\rm rel}^{\mathbb{Y'}}(\Phi)=H(\frac{9}{16})-H(\frac{1+\sqrt{1-\lambda}}{2})+{\rm log}2$, which is equal to the right hand side of (\ref{eq rel}), $H(\sqrt{c_{\rm max}})-2S(\frac{J_\Phi}{2})+2{\rm log}2$. In this case, the phase damping channel satisfies the condition $|\la x|\Phi|i\ra|^2+|\la y'|\Phi|i\ra|^2=1-\sqrt{c_{\rm max}},\ i=0,1$, the sum of the relative entropy coherence of the phase damping channel $\Phi$ with respect to the two bases $\mathbb{X}$ and $\mathbb{Y'}$ reaches the lower bound of (\ref{eq rel}).

\section{The quantum uncertainty relation of unitary channels with respect to the $l_1$ norm of coherence}

In this section, we consider the quantum uncertainty relation of {\it unitary} channels with respect to the $l_1$ norm of coherence. First we need the following lemma in Ref. \cite{yuan}.

{\bf Lemma 2.} \label{lemma 2}\cite{yuan}
Suppose $\vec{a}$, $\vec{b}$ and $\vec{c}\in \mathbb{R}^3$ are three-dimensional
nonzero vectors. We denote $\alpha$, $\beta$, and $\gamma$ to be the angles between
$\vec{a}$, $\vec{b}$; $\vec{b}$, $\vec{c}$; $\vec{c}$, $\vec{a}$ respectively, that is, $|\la \vec{a}| \vec{b}\ra|^2=\cos^2 \frac{\alpha}{2}$, $|\la \vec{b}| \vec{c}\ra|^2=\cos^2 \frac{\beta}{2}$, $|\la \vec{a}| \vec{c}\ra|^2=\cos^2 \frac{\gamma}{2}$ with $\alpha$, $\beta$, $\gamma\in[0,\pi]$. Then we have
\begin{eqnarray}
\sin\alpha+ \sin\beta \geqslant \sin\gamma.
\end{eqnarray}

Now we are ready to get the quantum uncertainty relation of unitary channels with respect to the $l_1$ norm of coherence.

{\bf Theorem 2.} For any given unitary channel $\Phi$ and any two measurement bases $\mathbb{X}=\{|x\ra,\ |x_\perp\ra\}$ and $\mathbb{Z}=\{|z\ra,\ |z_\perp\ra\}$, the quantum uncertainty relation of unitary channel with respect to the $l_1$ norm of coherence is
\begin{equation}\label{eq l1} C_{l_1}^\mathbb{X}(\Phi)+C_{l_1}^\mathbb{Z}(\Phi)\geqslant4\sqrt {c_{\rm max}(1-c_{\rm max})}+2.\end{equation}

{\bf Proof.}
Without loss of generality, we suppose the maximum overlap between these two bases $\mathbb{X}$ and $\mathbb{Z}$ is $c_{\rm max}=|\la x|z\ra|^2=\cos^2 \frac{\gamma}{2}$.
For any given unitary channel $\Phi$, suppose $\Phi(\cdot)=M(\cdot) M^\dagger$ with $M$ any unitary operator.
Under the orthonormal basis $\mathbb{X}=\{|x\ra,\ |x_\perp\ra\}$, the Choi matrix $J_\Phi$  is \begin{align}\begin{split}J_\Phi&=\begin{pmatrix}M|0\ra\la0|M^\dagger&M|0\ra\la1|M^\dagger\\M|1\ra\la0|M^\dagger&M|1\ra\la1|M^\dagger\end{pmatrix}\\
&=\begin{pmatrix}\la x|M|0\ra\la0|M^\dagger|x\ra&\la x|M|0\ra\la0|M^\dagger|x_\perp\ra&\la x|M|0\ra\la1|M^\dagger|x\ra&\la x|M|0\ra\la1|M^\dagger|x_\perp\ra\\
\la x_\perp|M|0\ra\la0|M^\dagger|x\ra&\la x_\perp|M|0\ra\la0|M^\dagger|x_\perp\ra&\la x_\perp|M|0\ra\la1|M^\dagger|x\ra&\la x_\perp|M|0\ra\la1|M^\dagger|x_\perp\ra\\
\la x|M|1\ra\la0|M^\dagger|x\ra&\la x|M|1\ra\la0|M^\dagger|x_\perp\ra&\la x|M|1\ra\la1|M^\dagger|x\ra&\la x|M|1\ra\la1|M^\dagger|x_\perp\ra\\
\la x_\perp|M|1\ra\la0|M^\dagger|x\ra&\la x_\perp|M|1\ra\la0|M^\dagger|x_\perp\ra&\la x_\perp|M|1\ra\la1|M^\dagger|x\ra&\la x_\perp|M|1\ra\la1|M^\dagger|x_\perp\ra\end{pmatrix}.\end{split}\end{align}
So the $l_1$ norm coherence of the unitary channel $\Phi$ is
\begin{align}\begin{split}C_{l_1}^\mathbb{X}(\Phi)=&|\la x|M|0\ra\la0|M^\dagger|x_\perp\ra|+|\la x|M|1\ra\la1|M^\dagger|x_\perp\ra|\\
&+|\la x|M|0\ra\la1|M^\dagger|x\ra|+|\la x|M|0\ra\la1|M^\dagger|x_\perp\ra|\\
&+|\la x_\perp|M|0\ra\la0|M^\dagger|x\ra|+|\la x_\perp|M|0\ra\la1|M^\dagger|x\ra|.
\end{split}\end{align}
Combining the condition $M^\dagger M=I$, we have $\la x|M|0\ra\la1|M^\dagger|x\ra=-\la x_\perp|M|0\ra\la1|M^\dagger|x_\perp\ra$ and $\la x|M|1\ra\la0|M^\dagger|x_\perp\ra=-\la x_\perp|M|1\ra\la0|M^\dagger|x_\perp\ra$. Multiplying the two equations, we can get $|\la x|M|0\ra|^2|\la x|M|1\ra|^2=|\la x_\perp|M|0\ra|^2|\la x_\perp|M|1\ra|^2$, i.e. $|\la x|M|0\ra|^2+|\la x|M|1\ra|^2=1$.
Suppose $|\la x|M|0|\ra|^2=\cos^2 \frac{\alpha}{2}$, then the $l_1$ norm of coherence of unitary channel under the orthonormal basis $\mathbb{X}$ is
\begin{align}
\begin{split}C_{l_1}^\mathbb{X}(\Phi)
=4\rm{cos}\frac{\alpha}{2}\rm{sin}\frac{\alpha}{2}+1
=2\rm{sin}\alpha+1.
\end{split}\end{align}
Similarly, under the orthonormal basis $\mathbb{Z}$, we suppose $|\la z|M|0|\ra|^2=\cos^2 \frac{\beta}{2}$,
then the $l_1$ norm of coherence of unitary channel under the orthonormal basis $\mathbb{Z}$ is $C_{l_1}^\mathbb{Z}(\Phi)=2\rm{sin}\beta+1$.
Therefore, the quantum uncertainty relation of the $l_1$ norm coherence of the unitary channel is
\begin{align}\begin{split}C_{l_1}^\mathbb{X}(\Phi)+C_{l_1}^\mathbb{Z}(\Phi)&=2(\rm{sin}\alpha+\rm{sin}\beta)+2\\
&\geqslant2\rm{sin}\gamma+2\\
&=4\sqrt {c_{\rm max}(1-c_{\rm max})}+2,\end{split}\end{align}
by Lemma 2.\hfill$\square$

The quantum uncertainty relation in Theorem 2 gives rise to the constraints of quantum coherence of unitary operations  in terms of $l_1$ norm of coherence under two orthonormal bases. This quantum uncertainty relation just relies on the incompatibility of two orthonormal bases and thus independent of the quantum channel. Now we illustrate the quantum uncertainty relation by two specific examples.

{\bf Example 4.} Now let us consider the unitary channel $\Phi$ with Kraus operator $E= \cos \alpha |0\rangle\langle 0| -\sin \alpha |0\rangle\langle 1|+ \sin \alpha |1\rangle\langle 0| +\cos \alpha |1\rangle\langle 1|$.
Given two measurement bases $\mathbb{X}=\{|0\ra,\ |1\ra\}$ and $\mathbb{Z}=\{|+\ra,\ |-\ra\}$ with $c_{\rm max}=1/2$. Then the Choi matrix of this unitary channel under the two orthonormal bases $\mathbb{X}$ and $\mathbb{Z}$ are
\begin{equation}
J_\Phi^{\mathbb{X}}=
\begin{pmatrix}
\cos^2\alpha&\sin\alpha\cos\alpha&-\sin\alpha\cos\alpha&\cos^2\alpha\\
\sin\alpha\cos\alpha&\sin^2\alpha&-\sin^2\alpha&\sin\alpha\cos\alpha\\
-\sin\alpha\cos\alpha&-\sin^2\alpha&\sin^2\alpha&-\sin\alpha\cos\alpha\\
\cos^2\alpha&\sin\alpha\cos\alpha&-\sin\alpha\cos\alpha&\cos^2\alpha
\end{pmatrix},
\end{equation}
and
\begin{equation}
J_\Phi^{\mathbb{Z}}=
\begin{pmatrix}
\frac{1}{2}(1+2\sin\alpha\cos\alpha)&\frac{1}{2}(\cos^2\alpha-\sin^2\alpha)&\frac{1}{2}(\cos^2\alpha-\sin^2\alpha)&\frac{1}{2}(-1-2\cos\alpha\sin\alpha)\\
\frac{1}{2}(\cos^2\alpha-\sin^2\alpha)&\frac{1}{2}(1-2\sin\alpha\cos\alpha)&\frac{1}{2}(1-2\sin\alpha\cos\alpha)&\frac{1}{2}(\sin^2\alpha-\cos^2\alpha)\\
\frac{1}{2}(\cos^2\alpha-\sin^2\alpha)&\frac{1}{2}(1-2\sin\alpha\cos\alpha)&\frac{1}{2}(1-2\sin\alpha\cos\alpha)&\frac{1}{2}(\sin^2\alpha-\cos^2\alpha)\\
\frac{1}{2}(-1-2\sin\alpha\cos\alpha)&\frac{1}{2}(\sin^2\alpha-\cos^2\alpha)&\frac{1}{2}(\sin^2\alpha-\cos^2\alpha)&\frac{1}{2}(1+2\sin\alpha\cos\alpha)
\end{pmatrix},
\end{equation}
respectively.
Therefore the $l_1$ norm of coherence with respect to these two measurement bases are
\begin{equation}
C_{l_1}^\mathbb{X}(\Phi)=2|\sin2\alpha|+1,\ {\text {and}}\ \  C_{l_1}^\mathbb{Z}(\Phi)=2|\cos2\alpha|+1,
\end{equation}
respectively.
Thus by Theorem 2, we get the lower bound of the quantum uncertainty relation of the unitary channel with respect to the $l_1$ norm of coherence is a constant 4.
While the sum of the quantum coherence is exactly $C_{l_1}^\mathbb{X}(\Phi)+C_{l_1}^\mathbb{Z}(\Phi)=2(|\sin2\alpha|+|\cos2\alpha|)+2$. Let $m(\alpha)=2(|\sin2\alpha|+|\cos2\alpha|)+2$ to represent the sum of the coherence of $\Phi$ under two bases. The comparison between the sum of the $l_1$ norm of coherence and the lower bound $m2(\alpha)=4$ is shown in Fig. 2. The lower bound of the quantum uncertainty relation is saturated for this unitary channel if and only if $\alpha=\frac{k\pi}{4}$ for arbitrary integer $k$.

\begin{figure}[h]
\includegraphics{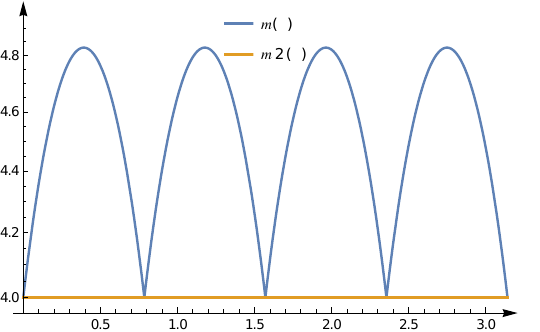}
\caption {The quantum uncertainty relations of unitary channel with respect to the $l_1$ norm of coherence in Example 4. $m(\alpha)=2(|\sin2\alpha|+|\cos\alpha|)+2$ shows the sum of the coherence of the channel under the measurement bases $\mathbb{X}$ and $\mathbb{Z}$. $m2(\alpha)=4$ is the lower bound of the quantum uncertainty relation by the right hand side of inequality (\ref{eq l1}).}
\end{figure}

{\bf Example 5.} Now we consider the unitary channel $\Phi$ with Kraus operator $\sigma_x=|1\rangle\langle0|+|0\rangle\langle1|$, $\Phi(\cdot)=\sigma_x(\cdot)\sigma_x$. Given two measurement bases $\mathbb{X}=\{|0\ra,\ |1\ra\}$ and $\mathbb{Z}=\{|+\ra,\ |-\ra\}$ with $c_{\rm max}=1/2$. Then the Choi matrices of the quantum channel $\Phi$ under these two orthonormal bases are
\begin{equation}J_\Phi^{\mathbb{X}}=\begin{pmatrix}0&0&0&0\\0&1&1&0\\0&1&1&0\\0&0&0&0\end{pmatrix}\ {\rm and}\ J_\Phi^\mathbb{Z}=\begin{pmatrix}\frac{1}{2}&-\frac{1}{2}&\frac{1}{2}&\frac{1}{2}\\-\frac{1}{2}&\frac{1}{2}&-\frac{1}{2}&-\frac{1}{2}\\
\frac{1}{2}&-\frac{1}{2}&\frac{1}{2}&\frac{1}{2}\\\frac{1}{2}&-\frac{1}{2}&\frac{1}{2}&\frac{1}{2}\end{pmatrix},\end{equation}
respectively. By calculation we get the sum of the $l_1$ norm of coherence is $C_{l_1}^\mathbb{X}(\Phi)+C_{l_1}^\mathbb{Z}(\Phi)=4$, which reaches exactly the lower bound of the quantum uncertainty relation in Theorem 2.

\section{Conclusion}
In this paper, we investigated the quantum uncertainty relation of quantum channel in qubit systems. For any two measurement bases, we derived the quantum uncertainty re- lations for quantum channels with respect to the relative entropy of coherence, and the quantum uncertainty relation for unitary channels with respect to the $l_1$ norm of coherence. These quantum uncertainty relations show the constraints of quantum coherence under dif- ferent measurement bases. Some examples are given in detail. Since quantum channels transform initial quantum states to certain final quantum states, they transmit quantum in- formation used in quantum tasks. Our quantum uncertainty relations for quantum channels give the constraints of the quantum coherence of quantum channels in different measurement bases. As the roles played by the uncertainty relations for quantum states in quantum key distribution, the uncertainty relations for quantum channels have also potential applications in related quantum information processing.

\bigskip
\noindent{\bf Acknowledgments} We thank the anonymous referees for useful suggestions and comments.
This work is supported by the National Natural Science Foundation of China under grant No. 12171044 and No. 12075159, and the specific research fund of the Innovation Platform for Academicians of Hainan Province.

\end{document}